\documentstyle[multicol,aps,prl]{revtex}
\newcommand{\tr}{{\rm tr}}
\newcommand{\ri}{{\rm i}}

\begin{document}

\title{Long--lived Quantum Coherence between Macroscopically Distinct States
in Superradiance}
\author{Daniel Braun$^{(1)}$, Petr A.~Braun$^{(1,2)}$ and Fritz Haake$^{(1)}$}
\address{$^{(1)}$ FB7, Universit\"at--GHS Essen, 45\,117 Essen,
Germany\\
$^{(2)}$ Department of Theoretical Physics, Institute of Physics,
Saint-Petersburg University, Saint-Petersburg 198904, Russia}

\maketitle
\begin{abstract}
The dephasing influence of a dissipative environment reduces linear
superpositions of macroscopically distinct  
quantum states (sometimes also called Schr\"odinger cat states) usually
{\em almost immediately} to a statistical mixture. This process is called
decoherence. 
Couplings to the environment with a certain symmetry can lead to {\em slow}
decoherence. 
In this Letter we show that the collective coupling of a large number of
two--level atoms to an electromagnetic field mode in a cavity that leads to
the phenomena of superradiance has such a symmetry, at least approximately. We 
construct superpositions of macroscopically distinct quantum states 
decohering only on a {\em classical} time scale and propose an experiment in
which the extraordinarily slow decoherence should be observable.
\end{abstract}

\begin{multicols}{2}
A Schr\"odinger cat state is a superposition of
two quantum states that differ on a
macroscopic scale. While
commonplace in the microscopic world, superpositions of macroscopically
distinct states have never been observed.
In other words, only probabilities and never
probability amplitudes are met with macroscopically. Our
understanding of why this is so has evolved considerably with the
development of the quantum mechanics of dissipative systems
\cite{Feynman63,Weiss93}: Dissipation due to the coupling to an
environment with a large number of degrees of freedom causes a
superposition $|\psi\rangle=c_1|\psi_1\rangle+c_2|\psi_2\rangle$ to
decohere very rapidly
towards the mixture
$\rho=|c_1|^2|\psi_1\rangle\langle\psi_1|+
|c_2|^2|\psi_2\rangle\langle\psi_2|$. 
``Rapidly'' means on a time scale much shorter than that on which the
mixture changes thereafter provided, of course, that $|\psi_1\rangle$
and $|\psi_2\rangle$ are macroscopically distinct. Recent
experiments of Haroche {\em et al.} \cite{Brune96,Haroche98} begin to
validate 
this picture by controlling the difference between two states put to a
superposition and following their decoherence. A general rule for the
ratio of time scales for decoherence ($T_{dec}$) and subsequent
near--classical motion ($T_{class}$) has become popular for the case
when one can associate a classical distance $D$ in phase space with the
pair of states $|\psi_1\rangle$, $|\psi_2\rangle$, $T_{dec}/T_{class}
=(\hbar/D)^p$ with some positive power $p$
\cite{Zurek82,Caldeira83,WallsMilburn,Haake87,Zukowski,Strunz97}. The
phantastic smallness of this 
factor for macroscopic  
values of $D$ makes quantum states of the Schr\"odinger cat type so
unfamiliar in the 
macroscopic world.

The situation is different, however, if the coupling to the environment has
a symmetry in the following sense. Suppose that in a Hamiltonian
$H_{int}=\hat{A}f(\{ {\bf x},{\bf p} \})$ a  
``coupling agent'' $\hat{A}$ couples the system to the
environment; the position and momentum operators ${\bf x}$ and
${\bf p}$ represent environmental degrees of freedom. With symmetry of the
coupling we mean that $\hat{A}$ has at least one degenerate eigenvalue,
i.e.~there are 
at least two linearly independent states $|\Phi_1\rangle$ and
$|\Phi_2\rangle$  such that $\hat{A}|\Phi_1\rangle=a|\Phi_1\rangle$ and
$\hat{A}|\Phi_2\rangle=a|\Phi_2\rangle$. It has been pointed out by Zurek that 
rapid decoherence arises only between subspaces pertaining to different
eigenvalues of $\hat{A}$ \cite{Zurek82}, whereas the coherence of any
syperposition within the subspace pertaining to the same degenerate
eigenvalue is not affected by the environment. The linear combination
$|\Phi\rangle 
=c_1|\Phi_1\rangle+c_2|\Phi_2\rangle$ does therefore {\em not} show accelerated
decoherence, even if $|\Phi_1\rangle$ and $|\Phi_2\rangle$ are
macroscopically distinct. Loosely speaking, the environment cannot
distinguish in which of the two states the system is. The state $|\Phi\rangle$
remains in principle a pure state for ever if the system dynamics is
restricted to the subspace pertaining to the degenerate eigenvalue $a$ of
the coupling agent. In
the context of Markovian master equations this
was recently proven rigorously by Lidar {\em et al.}; and an application to 
decoherence free quantum computing was proposed \cite{Lidar98}. As we shall
see below, longevity may even survive weak symmetry breaking.

In the following we briefly recall the quantum optical system
that gives rise to superradiance. We show that the coupling to the
environment has a symmetry in the above sense and identify coherent states
as approximate eigenstates of the coupling agents. We calculate the
decoherence rates for superpositions of two coherent states and identify the
superpositions with long--lived quantum coherence. In the end we propose an
experimental realization.  

The system considered is well known from superradiance
\cite{Haroche82} where many identical two--level atoms in a cavity
radiate collectively. We formally assign to each two--level atom a
spin--$\frac{1}{2}$ operator whose $z$--component tells us whether the
atom is in its lower or upper state. The collection of atoms is
described by a Bloch vector ${\bf J}$, the sum of all
spin--$\frac{1}{2}$ operators and thus formally an angular momentum
with maximal amplitude $j=N/2$, which is huge if the number $N$ of atoms
is. The dynamics conserves the square of ${\bf J}$. The Hilbert space
therefore decomposes into $(2j+1)$--dimensional subspaces with ${\bf
J}^2=j(j+1)$, of which we shall always imagine the one
with the maximal $j=N/2$ selected.  The atoms
are coupled to the electromagnetic field in the cavity by an interaction
of the form
\begin{equation}
\label{Hint}
 H_{int}=\hbar g(J_+a+J_-a^\dagger)\,,
\end{equation}
where $J_\pm=J_x\pm iJ_y$ are
raising and lowering operators of the angular momentum while $a^\dagger$
and $a$ are creation and annihilation operators for an electromagnetic
mode of the cavity in resonance with the atomic transition. That mode is
itself damped due to its coupling to the continuum of modes of the
electromagnetic field outside the cavity. The coupling constant $g$ is a
Rabi frequency for the coupling of a single atom with the mode vacuum.
Dissipation ultimately arises by the non--ideal mirrors of the cavity
which allow photons to leak out and lead to an effectively
permanent energy loss of the atomic system.

Under reasonable simplifying assumptions (the most important are: (i)
good time scale separation between the reservoir and the system composed of  
atoms and resonator mode, (ii) environment temperature in the range $\hbar
\kappa\ll k_BT\ll \hbar \omega_0$ where $\kappa$ is the damping rate of
the mode amplitude, and (iii) $\kappa\gg g\sqrt{N}$ to suppress
damped oscillations of the Bloch vector) the dynamics of ${\bf J}$ is
well described by a master equation for the reduced atomic density
operator $\rho$ \cite{Bonifacio71.1},
\begin{equation}
\label{eq:rhotd}
\frac{d}{d\tau}\rho(\tau)=\frac{1}{2j}([J_-,\rho(\tau)
J_+]+[J_-\rho(\tau), J_+])\equiv\Lambda\rho(\tau)\,.
\end{equation}
The dimensionless time $\tau$ is in units of the
inverse classical damping rate, $\tau=t/T_{class}$. 
Note that the rate of transitions of a {\em single} atom from its upper
state to the 
ground state is $g^2/\kappa$ while $Ng^2/\kappa$ is the rate for the
{\em collective} motion of all atoms. It is the latter rate which gives the
classical time scale $T_{class}=\kappa/(Ng^2)$ of the superradiant
dynamics. Moreover, photons extracted from the initial atomic excitation
become available for detection outside the cavity on the time scale
$T_{class}$ which is $N$ times shorter than for non--collective
radiation. 

As $j$ is connected to the number of
atoms by $j=N/2$ it is obvious
that the system becomes macroscopic for $j\to \infty$. This limit
corresponds in fact to the classical limit, as can be seen more formally 
from the fact that the classical
phase space of the problem is the unit sphere ${\bf J}^2=
j(j+1)=const.$ \cite{Haake91}. Since it contains $2j+1$ states in the
representation selected   we may think of $\hbar$ as 
represented by $1/j$. The smallness of this parameter compared to $1$
decides how close we are to the classical limit $\hbar\to 0$. 
In the experiments $N$ was typically of the order of $10^5$
\cite{Haroche76}. One 
might thus prefer to speak of mesoscopic rather than macroscopic angular
momentum states.

Classically ${\bf J}$ behaves
like an overdamped pendulum: If we specify the orientation of
${\bf J}/j$ by a polar angle $\theta$ and an azimutal one $\varphi$, the
classical limit of (\ref{eq:rhotd}) yields $d\theta/d\tau=\sin\theta$ and
$\varphi=\mbox{\rm const.}$. This proves that $\tau$ is indeed the time in
units 
of the classical time scale. 
An exact solution of (\ref{eq:rhotd}) was already
obtained in \cite{Bonifacio71.1}; a systematic semiclassical treatment can
be found in \cite{Braun98.1}. But most importantly, it has been experimentally
verified that  
the superradiance master equation (\ref{eq:rhotd}) correctly
describes the radiation by identical atoms resonantly coupled to a leaky
resonator mode \cite{Haroche76}.

The states of the system that correspond most closely to classical
states are the so--called angular momentum coherent states
$|\gamma\rangle$ \cite{Glauber72,Arecchietal}. They correspond to a
classical angular momentum pointing in the direction given by $\theta$
and $\varphi$ with minimal uncertainty $\sim 1/j$. The complex amplitude
$\gamma$ 
is connected with $\theta$ and $\varphi$ via
$\gamma=\tan(\theta/2)e^{i\varphi}$. In terms of eigenstates
$|j,m\rangle$ of ${\bf J}^2$ and $J_z$ (with the respective eigenvalues
$j(j+1)$ and $m$) one has the expansion
\begin{equation}
\label{cs}
|\gamma\rangle=(1+\gamma\gamma^*)^{-j}\sum_{m=-j}^j\gamma^{j-m}\sqrt{2j\choose
j-m}|j,m\rangle\,.
\end{equation}
The angular momentum coherent states are approximate
eigenstates of $J_-$ in 
the sense that the angle between $J_-|\gamma\rangle$ and $|\gamma\rangle$
is of the order of $1/\sqrt{j}$. Indeed, if we define this angle $\alpha$
for real $\gamma$ by $|\langle \gamma|J_-|\gamma\rangle|^2=\langle \gamma|
\gamma\rangle\langle \gamma|J_+J_-|\gamma\rangle
\cos^2\alpha$ one
easily shows that 
\begin{eqnarray} \label{al}
\cos^2\alpha&=&\frac{\sin^2\theta}{\sin^2\theta+\frac{2}{j}\cos^4\frac{\theta}{2}}\\
&=&1-\frac{1}{j}\frac{1}{2\gamma^2}+{\cal O}(\frac{1}{j^2})\mbox{
for }\sin\theta\ne 0\label{al2}\,.
\end{eqnarray}
The corresponding approximate eigenvalue is given by
$J_-|\gamma\rangle\simeq j\sin\theta e^{-\ri \varphi}|\gamma\rangle$ and 
immediately reveals a fundamental symmetry: Since
$\sin\theta_1=\sin\theta_2$ for 
$\theta_2=\pi-\sin\theta_1$, two different coherent states
$|\gamma_1\rangle$ and 
$|\gamma_2\rangle$ can have the same approximate eigenvalue.

Consider now a Schr\"odinger cat state composed of two coherent states,
\begin{equation} \label{cat}
|\Phi\rangle=c_1|\gamma_1\rangle+c_2|\gamma_2\rangle\,,
\end{equation}
where $c_1$ and $c_2$ are properly normalized but otherwise arbitrary
complex coefficients.
We will show that $ |\Phi\rangle$ in general experiences decoherence
on the scale $\tau\sim 
\frac{1}{j}$ which is shorter by the factor $1/j$ than the classical time  
scale. Our central prediction is,
however, that such
{\em accelerated decoherence is absent for Schr\"odinger cat states with  
$\gamma_1\gamma_2^*=1$}. Such two states  
correspond to two
classical angular momenta arranged in the plane
$\varphi=\varphi_1=\varphi_2$ symmetrically with respect to the equator
$\theta=\pi/2$ and therefore profit from the mentioned symmetry in the
coupling, i.e.~$\sin\theta_1=\sin\theta_2$. 
A particular such
state is the ``antipodal polar'' one with $\gamma_1=0$ and
$\gamma_2=\infty$ alias $\theta_1=0$ and $\theta_2=\pi$; in superradiance
parlance, it is the superposition of ``all up'' and ``all down``.\\

To prove the longevity of the exceptional Schr\"odinger cat states we need
to look at 
the initial density matrix $\rho(0)=|c_1|^2|\gamma_1\rangle\langle
\gamma_1|+c_1c_2^*|\gamma_1\rangle\langle \gamma_2|+\ldots$. Since the
evolution 
equation (\ref{eq:rhotd}) of $\rho(\tau)=\exp(\Lambda\tau)\rho(0)$ is
linear it suffices to discuss the fate of
$\exp(\Lambda\tau)(|\gamma\rangle\langle \gamma'|)\equiv
\rho(\gamma,\gamma',\tau)$ where $\gamma$ and $\gamma'$ may the take
values $\gamma_1$ and $\gamma_2$. The relative weights of the four
$\rho(\gamma,\gamma',\tau)$ can be studied in terms of either one of the
norms
\begin{eqnarray}
\label{norms}
N_1(\gamma,\gamma'\tau)&=&
\rm{tr}\,\rho(\gamma,\gamma',\tau)\rho^\dagger(\gamma,\gamma',\tau)
\nonumber\,,\\
N_2(\gamma,\gamma'\tau)&=&
\sum_{m_1,m_2=-j}^j|\langle j,m_1|\rho(\gamma,\gamma',\tau)|j,m_2\rangle|
\,.
\end{eqnarray}
We shall see presently that these norms evolve on the classical time
scale ${\cal O}(1)$ for $\gamma=\gamma'$ while $\gamma\ne\gamma'$ in
general yields decay on the shorter time scale $\sim 1/j$; it is only
the exceptional offdiagonal cases given above which decay on the
classical time scale.

The foregoing assertions are based on the following three analytical
results:

(i) The initial time derivative of $N_1(\gamma,\gamma',\tau)$ reads
\begin{eqnarray}
&&\frac{dN_1(\gamma_1,\gamma_2,\tau)}{d\tau}\Big|_{\tau=0}=-
\frac{1}{2}\left((1+\cos\theta_1)^2+(1+\cos\theta_2)^2\right)\nonumber \\&-&
j\left(\sin^2\theta_1+\sin^2\theta_2-2\cos(\varphi_2-\varphi_1)
\sin\theta_1\sin\theta_2\right)\label{der}\,.
\end{eqnarray}
The first term obviously describes evolution on the classical time scale
while the second, being proportional to $j$, can give rise to motion
on the shorter time scale $1/j$. The ``fast'' term vanishes for
$\gamma_1=\gamma_2$; while it is present in general for $\gamma_1\ne
\gamma_2$ we read off its absence for the exceptional cases
$\varphi_1=\varphi_2$, $\sin\theta_1=\sin\theta_2$; since $\sin \theta$ is
symmetrical about $\theta=\pi/2$ the long--lived Schr\"odinger cat states
are indeed 
revealed as superposition of coherent states oriented symmetrically with
respect 	
to the equatorial plane. The proof of (\ref{der}) is a back-of-the-envelope
calculation involving the expectation values $\langle
\gamma|J_{\pm}|\gamma\rangle=je^{\pm\ri\varphi}\sin\theta$ and $\langle
\gamma|J_+J_-|\gamma\rangle=|\langle
\gamma|J_-|\gamma\rangle|^2+2j\cos^4(\theta/2)$ in
$\frac{dN_1(\gamma,\gamma',\tau)}{d\tau}=\tr(\frac{d\rho}{d\tau}\rho^\dagger+
\rho\frac{d\rho^\dagger}{d\tau})$.

(ii)  For the polar antipodal Schr\"odinger cat state
$|\Phi\rangle=\frac{1}{\sqrt{2}}(|j,j\rangle+|j,-j\rangle)$ corresponding to
$\gamma_1=0$ and $\gamma_2=\infty$ or $\theta_1=0$ and $\theta_2=\pi$, the
time--dependent norms are easily found exactly. The master equation
(\ref{eq:rhotd}) yields a single differential equation for the matrix
element $\langle j,j |\rho(\tau)|j,-j\rangle$, $\frac{d}{d\tau}\langle j,j|\rho|j,-j\rangle=-\langle j,j |\rho(\tau)|j,-j\rangle$.
The norm of the offdiagonal parts therefore decays as 
$N_1(0,\infty,\tau)=N_1(\infty,0,\tau)=e^{-2\tau}$, i.e.~on
classical time scale!  The same conclusion can be
drawn from the second norm which yields
$N_2(0,\infty,\tau)=N_2(\infty,0,\tau)=e^{-\tau}$. The polar antipodal cat
is therefore definitely long--lived.

(iii) Inasmuch as the initial slope
$\frac{d}{d\tau}N_1(\gamma,\gamma',\tau)\left.\right|_{\tau=0}$
could
in principle be deceptive it is  desirable to follow the evolution for the
general state(\ref{cat}) to
positive times, at least to times of order $\tau\sim 1/j$.
For times $\tau$ with $0\le j\tau\ll 1$, a somewhat
involved semiclassical evaluation
of the norm $N_2(\tau)$ for $\varphi_1=\varphi_2$ (for simplicity we
restrict ourselves to $\varphi_1=0=\varphi_2$, i.e.~to real
$\gamma_1$,$\gamma_2$), which we shall present elsewhere, leads to
\begin{equation} \label{31}
\frac{N_2(\gamma_1,\gamma_2,\tau)}{N_2(\gamma_1,\gamma_2,0)}\approx
\exp\left(-2j\frac{(\gamma_1-\gamma_2)^2(1-\gamma_1\gamma_2)^2}
{((1+\gamma_1^2)(1+\gamma_2^2))^2}\tau\right)\,.
\end{equation}
The corrections are of relative order $1/j$. This means accelerated
decoherence as long as $\gamma_1\ne\gamma_2$ 
and $\gamma_1\gamma_2\ne 1$. If, however, $\gamma_1\gamma_2=1$ the next
order in $1/j$ shows that
\begin{eqnarray} \label{32}
\frac{N_2(\gamma_1,\frac{1}{\gamma_1},\tau)}{N_2(\gamma_1,\frac{1}
{\gamma_1},0)}=&&\exp\Big(-\left(\frac{\gamma_1^2-1}{\gamma_1^2+1}\right)^2
\tau\\
&&-\frac{3\gamma_1^8-3\gamma_1^6+4\gamma_1^4-3\gamma_1^2+3}
{2(\gamma_1^2+1)^4}\tau^2\Big)\nonumber\,.
\end{eqnarray}
This expression is correct up to and including order $(j\tau)^2$. Obviously,  
accelerated
decoherence is absent in this case. Indeed, a
pure coherent state $\gamma_1=\gamma_2=\gamma$ leads, in linear order in
$j\tau$, to
almost the same decay,
\begin{equation} \label{33}
\frac{N_2(\gamma,\gamma,\tau)}{N_2(\gamma,\gamma,0)}=
\exp\left(-\gamma^4\left(\frac{\gamma^2-1}{\gamma^2+1}\right)^2\tau\right)\,.
\end{equation}

The long--lived coherences described above
should be experimentally observable. We now propose a scheme for
the preparation of the special long--lived Schr\"odinger cat states. It is
based on the result by Agarwal {\it et al} \cite{Agarwal97} that a strongly
detuned cavity 
(detuning $\delta$ between mode frequency and atomic frequency $\gg\kappa$)
leads to an effectively unitary evolution
$\frac{d\rho}{dt}=-\frac{i}{\hbar}[H_{eff},\rho]$ with a non--linear
Hamiltonian 
\begin{equation} \label{heff}
H_{eff}=\hbar \eta J_+J_-\mbox{ with }\eta=\frac{g^2\delta}{\kappa^2+\delta^2}\,.
\end{equation}
Due to the nonlinearity a coherent state $|\theta,\varphi\rangle$ evolves
into a superposition of several coherent states. At time
$t=\frac{\pi}{m\eta}$ where $m$ is an even integer, the state consists of
$m$ components $|\theta,\varphi+\pi\frac{2q-N+1}{m}\rangle$,
$q=0,1,\ldots,m-1$. In particular, after the time $t=\frac{\pi}{2\eta}$ the
coherent state has evolved into the superposition
\begin{equation} \label{sup}
\frac{e^{-\ri (N-\frac{1}{2})\frac{\pi}{2}}}{\sqrt{2}}\left(|\theta,\varphi-\pi\frac{N-1}{2}\rangle-\ri|\theta,\varphi-\pi\frac{N-3}{2}\rangle
\right)\,,
\end{equation} 
i.e.~the azimuths of the two components differ by $\pi$.
The preparation of the special long--lived Schr\"odinger cat states involves
three steps. During the whole preparation the dissipation due to the
superradiance damping can be effectively turned off by the described strong
detuning of the cavity.
 
One starts with the mode in the vacuum state and all atoms in
their ground state and applies a resonant coherent laser pulse to bring
the atoms in a coherent state $|\theta,\varphi\rangle$ with $j=N/2$.
Second, one lets the atoms evolve freely in the detuned cavity for the time
$t=\frac{\pi}{2\eta}$ and thus  ``dispersively'' turns the
coherent state into a superposition of two such,
$|\theta,\varphi'\rangle$ and $|\theta,\varphi'+\pi\rangle$, as 
described above. Finally, a resonant coherent $\pi/2$ pulse brings the
superposition to the 
desired orientation symmetric to the equator, by rotation through the
angle $\pi/2$ about an axis perpendicular to the plane defined by the
directions of the two coherent states produced in step two. The angle
$\theta$ which was the same for both components after step two becomes the
angle with respect to the equator.

Presently
used techniques for preparing very cold two-level atoms in a
superposition of ``up'' and ``down'' and transporting them inside
high-$Q$ resonators work atom by atom \cite{Brune96,Haroche98}; they might
be modified so as to 
place collections of several such atoms into a resonator, thus producing
the long--lived Schr\"odinger cat state.

We have enjoyed and profited from discussions with Girish Agarwal, Serge  
Haroche, Walter Strunz, Dan Walls, and Peter Zoller. Financial support by the Sonderforschungsbereich
``Unordnung
und gro{\ss}e Fluktuationen'' of the Deutsche Forschungsgemeinschaft is
gratefully acknowledged. An anonymous referee deserves a thankyou for
bringing Ref.\cite{Lidar98} to our attention.

\end{multicols}

\end{document}